\documentclass[prd,aps,showpacs,epsf,floats,10pt]{revtex4}
\usepackage{amssymb}
\usepackage{amsfonts}
\usepackage{amsmath}

\input{tcilatex}
\begin{document}

\title{ON GROVER'S SEARCH ALGORITHM FROM A QUANTUM INFORMATION GEOMETRY
VIEWPOINT}
\author{Carlo Cafaro$^{1}$ and Stefano Mancini$^{2}$}
\affiliation{$^{1\text{, }2}$School of Science and Technology, Physics Division,
University of Camerino, I-62032 Camerino, Italy}

\begin{abstract}
We present an information geometric characterization of Grover's quantum
search algorithm. First, we quantify the notion of quantum
distinguishability between parametric density operators by means of the
Wigner-Yanase quantum information metric. We then show that the quantum
searching problem can be recast in an information geometric framework where
Grover's dynamics is characterized by a geodesic on the manifold of the
parametric density operators of pure quantum states constructed from the
continuous approximation of the parametric quantum output state in Grover's
algorithm. We also discuss possible deviations from Grover's algorithm
within this quantum information geometric setting.
\end{abstract}

\maketitle

\section{Introduction}

Information geometry is the application of differential geometric techniques
to the study of families of probabilities, both classical and quantum,
either parametric or nonparametric \cite{amari}. Much of the work in its
quantum version has been concentrated on manifolds of density operators for
both finite and infinite dimensional quantum systems. Classical and quantum
Fisher information metrics on statistical manifolds play a key-role in
information geometry, especially in the geometric and informational approach
to classical and quantum estimation theory, respectively \cite{amari}.
Applications of information geometric techniques appear in statistical
mechanics, in the study of the three-dimensional spherical model \cite%
{PRE2003, Physica A}, the Ising model on planar random graphs \cite{PRE2002,
Physica A} and the one-dimensional Potts model \cite{JPA, Physica A}.
Furthermore, the Fisher information plays a fundamental role in quantifying
Heisenberg's uncertainty principle from a statistical inference point of
view \cite{braunstein-caves}. Indeed, information geometric methods in
combination with probable inference techniques have been recently employed
to define the notion of distinguishability between probability distributions
of curved statistical manifolds underlying the information dynamics used to
provide a probabilistic description of physical systems when only partial
knowledge (incomplete information) about them is available \cite{carloIJTP,
carloMPLB, carloPA, carloPD} . This line of investigation is very
stimulating and appealing, especially considering that\ it is an old dream
that of viewing quantum mechanics as rooted in making statistical inferences
based on observed experimental data \cite{summhammer, wheeler}. Recent works
where information geometry and inference methods are used to investigate the
origin of fundamental theories as information geometric inferential theories
appear in \cite{caticha-cafaro, goyal, caticha}.

Very recently, we have observed that information theory has motivated
conventional differential geometric investigations of quantum state spaces.
For instance, Riemannian geometric tools have been used to characterize the
quantum gate complexity in quantum computing \cite{nielsen1, nielsen2,
brandt}. In \cite{nielsen1}, the problem of finding optimal quantum circuits
was recast as a differential geometric problem. Specifically, a Riemannian
metric on a manifold of multi-qubit unitary transformations is used to
define a metric distance between the identity operator and the desired
unitary operator representing the quantum computation. The metric distance
turns out to be equivalent to the number of quantum gates needed to
represent that unitary operator. It was then shown that finding optimal
quantum circuits is essentially equivalent to finding the shortest path
(geodesic) between two points in a certain curved geometry. However, the
number of investigations concerning the applications of quantum information
geometric methods to specific problems of quantum computing, such as quantum
estimation, quantum channel entropy, quantum noise and entanglement and
quantum algorithms is very limited. This is not unreasonable considering
that the systematic study of quantum information geometry has only started
no more than ten years ago \cite{amari}.

In this paper, we provide a simple application of quantum information
geometry to quantum computing. It is known that Grover's quantum search
algorithm \cite{grover} can be viewed as a geodesic path on the manifold of
Hilbert-space rays where the notion of quantum distinguishability is
quantified via the Fubini-Study metric, a gauge invariant metric on the
projective Hilbert space \cite{wadati, alvarez}. By observing that a
parametric quantum wavefunction induces in a natural manner a parametric
density operator and by considering its square root, we use the concept of
Wigner-Yanase quantum information metric. Such metric is one among many
versions of a so-called quantum Fisher information metric, a metric on
manifolds of density operators for both finite and infinite dimensional
quantum systems. We show in an explicit manner that the Wigner-Yanase metric
and the Fubini-Study metric differ by a factor of four when considering pure
state models. Finally, interpreting the Fubini-Study metric as a quantum
version of Fisher metric, we provide an information geometric
characterization of Grover's algorithm as a geodesic (shortest length curve)
in the parameter space characterizing the pure state model, the manifold of
the parametric density operators of pure quantum states. Finally, we discuss
few possible deviations (non-constant Fisher information function and/or
non-actuality) from Grover's algorithm within this quantum information
geometric framework.

The layout of the article is as follows. In Section II, we introduce the
Wigner-Yanase quantum information metric from both formal and heuristic
viewpoints. In Section\ III, we briefly discuss Grover's search algorithm in
quantum computing. In Section IV, we show that Grover's dynamics corresponds
to the shortest path from an information geometric point of view. In Section
V, we consider deviations from Grover's probability path where the
requirements of constant Fisher information function and/or actuality are
not simultaneously satisfied. Our concluding remarks appear in Section VI.

\section{On Information Geometry and Statistical Distinguishability}

In the classical setting, it is known that except for an overall
multiplicative constant the classical Fisher information metric is unique 
\cite{cencov, campbell}: it is the only monotone Riemannian metric with the
property of having its line element reduced under Markov morphisms
(stochastic maps). Said otherwise, there is essentially one classical
statistical distance quantifying the classical distinguishability between
two probability distributions. In the quantum setting, Riemannian metrics
are considered on the space of density matrices. The requirement that the
distance between density matrices expresses quantum statistical
distinguishability implies that this distance must decrease under
coarse-graining (stochastic maps). Unlike the classical case, it turns out
that there are infinitely many monotone Riemannian metrics satisfying this
requirement \cite{morozova, petz, grasselli}.

In this section we introduce the Wigner-Yanase quantum information metric
from both a formal and heuristic point of view.

\subsection{The Wigner-Yanase Quantum Information Metric: the Formal Approach%
}

Assume that $\mathcal{M}_{n}$ denotes the space of $\left( n\times n\right) $%
-complex matrices, $\mathcal{D}_{n}$ is the manifold of strictly positive
elements of $\mathcal{M}_{n}$ and $\mathcal{D}_{n}^{1}\subset $ $\mathcal{D}%
_{n}$ is the submanifold of density matrices,%
\begin{equation}
\mathcal{D}_{n}^{1}\overset{\text{def}}{=}\left\{ D\in \mathcal{D}%
_{n}:D=D^{\dagger }\text{and Tr}\left( D\right) =1\right\} \text{.}
\end{equation}%
The tangent space to $\mathcal{D}_{n}^{1}$ at $\rho $ is given by,%
\begin{equation}
T_{\rho }\mathcal{D}_{n}^{1}\overset{\text{def}}{=}\left\{ D\in \mathcal{M}%
_{n}:D=D^{\dagger }\text{and Tr}\left( D\right) =0\right\} \text{.}
\end{equation}%
Given $D_{1}$ and $D_{2}$ in $T_{\rho }\mathcal{D}_{n}^{1}$, the
Wigner-Yanase quantum information metric is defined as \cite{gibilisco},%
\begin{equation}
\left\langle D_{1}\text{, }D_{2}\right\rangle _{\rho \text{, }f_{\text{WY}}}%
\overset{\text{def}}{=}\text{Tr}\left[ D_{1}\cdot c_{\text{WY}}\left(
L_{\rho }\text{, }R_{\rho }\right) \left( D_{2}\right) \right] \text{,}
\label{wy-product}
\end{equation}%
where $L_{\rho }$ and $R_{\rho }$ are the left and right multiplication
operators, respectively, with%
\begin{equation}
L_{\rho }\left( D\right) \overset{\text{def}}{=}\rho D\text{ and, }R_{\rho
}\left( D\right) \overset{\text{def}}{=}D\rho \text{.}
\end{equation}%
The quantity $c_{\text{WY}}\left( L_{\rho }\text{, }R_{\rho }\right) $ in (%
\ref{wy-product}) denotes the so-called Chentsov-Morozova function for the
Wigner-Yanase information metric,%
\begin{equation}
c_{\text{WY}}\left( x\text{, }y\right) \overset{\text{def}}{=}\frac{1}{yf_{%
\text{WY}}\left( \frac{x}{y}\right) }=\frac{4}{\left( \sqrt{x}+\sqrt{y}%
\right) ^{2}}\text{ with }x\text{, }y>0\text{.}
\end{equation}%
The function $f:\left( 0\text{, }+\infty \right) \rightarrow 
\mathbb{R}
$ is an operator monotone function, that is $\forall n\in 
\mathbb{N}
$ and $\forall $ $M_{1}$, $M_{2}\in \mathcal{M}_{n}$ with $0\leq M_{1}\leq
M_{2}$, it must be $0\leq f\left( M_{1}\right) \leq f\left( M_{2}\right) $.
To evaluate $\left\langle D_{1}\text{, }D_{2}\right\rangle _{\rho \text{, }%
f_{\text{WY}}}$ in (\ref{wy-product}), observe that $T_{\rho }\mathcal{D}%
_{n}^{1}$ can be decomposed in two orthogonal subspaces $\left( T_{\rho }%
\mathcal{D}_{n}^{1}\right) ^{\text{c}}$ and $\left( T_{\rho }\mathcal{D}%
_{n}^{1}\right) ^{\text{o}}$,%
\begin{equation}
T_{\rho }\mathcal{D}_{n}^{1}=\left( T_{\rho }\mathcal{D}_{n}^{1}\right) ^{%
\text{c}}\oplus \left( T_{\rho }\mathcal{D}_{n}^{1}\right) ^{\text{o}}\text{,%
}
\end{equation}%
where,%
\begin{equation}
\left( T_{\rho }\mathcal{D}_{n}^{1}\right) ^{\text{c}}=\left\{ D\in T_{\rho }%
\mathcal{D}_{n}^{1}:\left[ D\text{, }\rho \right] =0\right\} \text{,}
\end{equation}%
and $\left( T_{\rho }\mathcal{D}_{n}^{1}\right) ^{\text{o}}$ is the
orthogonal complement of $\left( T_{\rho }\mathcal{D}_{n}^{1}\right) ^{\text{%
c}}$. Therefore, an arbitrary element $D^{\prime }$ of $\left( T_{\rho }%
\mathcal{D}_{n}^{1}\right) ^{\text{o}}$ can be written as $i\left[ \rho 
\text{, }D\right] $ with $D=D^{\dagger }$. Thus,%
\begin{equation}
\left\langle D^{\prime }\text{, }D^{\prime }\right\rangle _{\rho \text{, }f_{%
\text{WY}}}=\left\langle i\left[ \rho \text{, }D\right] \text{, }i\left[
\rho \text{, }D\right] \right\rangle _{\rho \text{, }f_{\text{WY}}}=\text{Tr}%
\left\{ i\left[ \rho \text{, }D\right] 4\left( L_{\rho }^{\frac{1}{2}%
}+R_{\rho }^{\frac{1}{2}}\right) ^{-2}\left( i\left[ \rho \text{, }D\right]
\right) \right\} =-4\text{Tr}\left( \left[ \rho ^{\frac{1}{2}}\text{, }D%
\right] ^{2}\right) \text{,}  \label{EX}
\end{equation}%
where we have made use of the following two relations,%
\begin{equation}
\left[ \rho \text{, }D\right] =\left( L_{\rho }-R_{\rho }\right) \left(
D\right) =\left( L_{\rho }^{\frac{1}{2}}+R_{\rho }^{\frac{1}{2}}\right)
\left( L_{\rho }^{\frac{1}{2}}-R_{\rho }^{\frac{1}{2}}\right) \left(
D\right) \text{, }
\end{equation}%
and,%
\begin{equation}
L_{\rho }^{\frac{1}{2}}\left( D\right) =\rho ^{\frac{1}{2}}D\text{, }R^{%
\frac{1}{2}}\left( D\right) =D\rho ^{\frac{1}{2}}\text{.}
\end{equation}%
We notice that $\left\langle D^{\prime }\text{, }D^{\prime }\right\rangle
_{\rho \text{, }f_{\text{WY}}}$ in (\ref{EX}) can be written as,%
\begin{equation}
I_{\text{WY}}\left( \rho \right) \equiv \left\langle D^{\prime }\text{, }%
D^{\prime }\right\rangle _{\rho \text{, }f_{\text{WY}}}=8I\left( \rho \text{%
, }D\right) \text{,}
\end{equation}%
where $I\left( \rho \text{, }D\right) $ defined as,%
\begin{equation}
I\left( \rho \text{, }D\right) \overset{\text{def}}{=}-\frac{1}{2}\text{Tr}%
\left( \left[ \rho ^{\frac{1}{2}}\text{, }D\right] ^{2}\right) \text{,}
\label{skew}
\end{equation}%
is the skew information introduced by Wigner and Yanase when studying
quantum measurement theory from an information-theoretic viewpoint \cite%
{wigner}.

\subsection{The Wigner-Yanase Quantum Information Metric: the Heuristic
Approach}

The classical Fisher information $I_{F}\left( p_{\theta }\right) $ of a
parametric probability density $p_{\theta }\left( x\right) $ is a
key-quantity in the statistical estimation theory and it is defined as,%
\begin{equation}
I_{F}\left( p_{\theta }\right) \overset{\text{def}}{=}\int dxp_{\theta
}\left( x\right) \left( \frac{\partial \log p_{\theta }\left( x\right) }{%
\partial \theta }\right) ^{2}\text{.}  \label{integro}
\end{equation}%
Observe that $I_{F}\left( p_{\theta }\right) $ can be rewritten in terms of
a fundamental quantity in quantum physics, the probability amplitude
(wavefunction) $\sqrt{p_{\theta }\left( x\right) }$,%
\begin{equation}
I_{F}\left( p_{\theta }\right) =4\int dx\left( \frac{\partial \sqrt{%
p_{\theta }\left( x\right) }}{\partial \theta }\right) ^{2}\text{.}
\label{integral}
\end{equation}%
When going from classical to quantum theory, a quantum version of the
classical Fisher information can be heuristically defined as follows: the
integral in (\ref{integral}) is replaced by the trace and the probability
densities $p_{\theta }$ by density operators $\rho _{\theta }$. Therefore $%
I_{F}\left( \rho _{\theta }\right) $, a quantum extension of $I_{F}\left(
p_{\theta }\right) $ to a family of parametric quantum operators $\rho
_{\theta }$, reads%
\begin{equation}
I_{F}\left( \rho _{\theta }\right) =4\text{Tr}\left( \frac{\partial \sqrt{%
\rho _{\theta }}}{\partial \theta }\right) ^{2}\text{.}  \label{qf}
\end{equation}%
Quantum generalizations of the very same classical Fisher information are
not unique. Indeed, two classically identical expressions (see (\ref{integro}%
)\ and (\ref{integral}), for instance) generally differ when they are
extended to a quantum setting. This difference is a manifestation of the
non-commutative nature of quantum mechanics and is reminiscent of the idea
of quantum discord \cite{luo}.

If $\rho _{\theta }$ in (\ref{qf}) satisfies the von Neumann equation,%
\begin{equation}
\frac{\partial \rho _{\theta }}{\partial \theta }+i\left[ T\text{, }\rho
_{\theta }\right] =0\text{,}
\end{equation}%
with $\rho _{0}=\rho $, $\theta \in 
\mathbb{R}
$ is a temporal parameter and $T$ is the generator of temporal shift. Then,%
\begin{equation}
\rho _{\theta }=e^{-i\theta T}\rho e^{i\theta T}\text{ and, }\frac{\partial 
\sqrt{\rho _{\theta }}}{\partial \theta }=ie^{-i\theta T}\left[ \rho ^{\frac{%
1}{2}}\text{, }T\right] e^{i\theta T}\text{.}  \label{relation}
\end{equation}%
Substituting the second equation in (\ref{relation}) into (\ref{qf}), $%
I_{F}\left( \rho _{\theta }\right) $ becomes independent of $\theta $ and
reads%
\begin{equation}
I_{F}\left( \rho _{\theta }\right) =8I\left( \rho \text{, }T\right) \text{,}
\end{equation}%
where $I\left( \rho \text{, }T\right) $ is the Wigner-Yanase skew
information defined in (\ref{skew}). Thus, the skew information is a
particular kind of quantum Fisher information $I_{F}\left( \rho _{\theta
}\right) $ known as the Wigner-Yanase quantum Fisher information 
\begin{equation}
I_{\text{WY}}\left( \rho _{\theta }\right) \equiv I_{F}\left( \rho _{\theta
}\right) =8I\left( \rho \text{, }T\right) \text{.}
\end{equation}%
When a density operator is parametrized by $n$-real parameters $\theta
\equiv \left( \theta ^{1}\text{,..., }\theta ^{n}\right) $ and the quantum
state $\rho _{\theta }$ satisfies the von Neumann equation, the set of
quantum states $\rho _{\theta }$ forms the quantum evolution submanifold of $%
\mathcal{D}_{n}^{1}$. The set of parameters $\theta $ can be regarded as a
local coordinate system on such submanifold endowed with a metric structure
defined by the information metric $g_{ij}^{\left( \text{WY}\right) }\left(
\theta \right) $,%
\begin{equation}
g_{ij}^{\left( \text{WY}\right) }\left( \theta \right) =4\text{Tr}\left[
\left( \partial _{i}\sqrt{\rho _{\theta }}\right) \left( \partial _{j}\sqrt{%
\rho _{\theta }}\right) \right] \text{, }\partial _{j}\equiv \frac{\partial 
}{\partial \theta ^{j}}\text{.}
\end{equation}%
Once the monotone Riemannian metric $g_{ij}^{\left( \text{WY}\right) }\left(
\theta \right) $ is explicitly obtained, the other geometric quantities
(Christoffel connection coefficients, scalar and sectional curvatures,
Riemannian curvature tensor, etc.) can be calculated from it in a purely
mathematical way, at least in principle. For instance, the geodesic equation
reads \cite{landau},%
\begin{equation}
D\dot{\theta}^{k}\overset{\text{def}}{=}\left( \frac{\partial \dot{\theta}%
^{k}}{\partial \theta ^{j}}+\Gamma _{ij}^{k}\left( \theta \right) \dot{\theta%
}^{i}\right) d\theta ^{j}=\frac{d^{2}\theta ^{k}\left( \tau \right) }{d\tau
^{2}}+\Gamma _{ij}^{k}\left( \theta \right) \frac{d\theta ^{i}}{d\tau }\frac{%
d\theta ^{j}}{d\tau }=0\text{,}  \label{GE}
\end{equation}%
where $D$ denotes the covariant derivative, $\dot{\theta}^{k}\overset{\text{%
def}}{=}\frac{d\theta ^{k}}{d\tau }$ and,%
\begin{equation}
\Gamma _{ij}^{k}\left( \theta \right) \overset{\text{def}}{=}\frac{1}{2}%
g^{kl}\left( \partial _{i}g_{lj}+\partial _{j}g_{il}-\partial
_{l}g_{ij}\right) \text{,}
\end{equation}%
are the Christoffel connection coefficients.

\section{On Grover's Search Algorithm}

In this section we briefly discuss what is considered a masterpiece of
quantum computational software \cite{mermin}, that is Grover's search
algorithm in quantum computing.

\subsection{Quantum Searching}

The search problem may be stated as follows: we wish to retrieve a certain
item satisfying a given condition assuming that it belongs to an unsorted
database (oracle) containing $N=$ $2^{n}$ items. One step is needed to
specify whether or not the examined item is the one fulfilling the given
condition. We assume that the selection of the item is aided by no sorting
on the database. In such a situation, the most efficient classical algorithm
to tackle the search problem requires the examination of the items in the
database one by one. Therefore, by means of a classical computer, the oracle
must be queried on average $\frac{N}{2}$ times ( $O\left( N\right) $
classical steps). However, by using the same amount of hardware as in the
classical case but by having the input and output in superpositions of
states, Grover has developed a quantum mechanical algorithm capable of
solving this search problem in about $\frac{\pi }{4}\sqrt{N}$ steps ( $%
O\left( \sqrt{N}\right) $ quantum mechanical steps) \cite{grover}. Although
this is not dramatic as the exponential quantum advantage achieved by Shor's
algorithm for factoring, the extremely wide applicability of searching
problems makes Grover's algorithm interesting and important. In particular,
Grover's algorithm gives a quadratic speed-up in the solution of $NP$%
-complete problems, which account for many of the important hard problems in
computer science. Indeed, it was shown in \cite{bennett} that relative to an
oracle chosen uniformly at random, with probability $1$, the class NP cannot
be solved on a quantum Turing machine in time $O\left( \sqrt{N}\right) $.
Drawing on this result, Grover pointed out in \cite{grover-d} that his
algorithm is optimal (i.e., the fastest), up to a multiplicative constant
factor, among all possible quantum algorithms. A detailed proof of Grover's
statement about the optimality of his quantum searching algorithm appears in 
\cite{boyer}. Finally, in \cite{zalka}, it is shown that for any number of
oracle lookups, Grover's algorithm is exactly (and not just asymptotically)
optimal.

What makes a quantum search algorithm more efficient than another? An
algorithm is an abstract mathematical concept, whereas it is useful to
consider how efficiently we can run an algorithm on a computer. Computer
scientists associate a cost with each step of the algorithm and with the
amount of memory required, embodying the idea that physical computers have a
finite size (memory) and work at a finite rate of elementary calculation
steps per unit time. This gives us a way to determine if one algorithm is
intrinsically faster than another. In general, it is very difficult to
formally prove that a certain algorithm is the "best" algorithm for a given
computational task. For instance, Shor's factoring algorithm \cite{shor97}
is only the "best known" algorithm for factoring; there is no proof that
something faster cannot be found in the future.

\subsection{Grover's Algorithm}

In what follows, we outline the construction of Grover's algorithm when
considering the $n$-qubit case ($N=2^{n}$ states). The step-$0$\ (the
initialization) of Grover's algorithm begins by using the Hadamard transform
to construct a uniform amplitude initial state which is an equal
superposition of all the orthonormal computational basis states $\left\{
\left\vert s\right\rangle \right\} $\ in the $N$-dimensional Hilbert space,%
\begin{equation}
\left\vert \bar{q}\right\rangle =\frac{1}{\sqrt{N}}\sum_{s=0}^{N-1}\left%
\vert s\right\rangle \text{.}  \label{input state}
\end{equation}%
Observe that the state (\ref{input state}) may be rewritten as,%
\begin{equation}
\left\vert \bar{q}\right\rangle =\frac{1}{\sqrt{N}}\sum_{s=0}^{N-1}\left%
\vert s\right\rangle =\sqrt{\frac{N-1}{N}}\sqrt{\frac{1}{N-1}}\sum_{s\neq
a}\left\vert s\right\rangle +\frac{1}{\sqrt{N}}\left\vert a\right\rangle
=\cos \left( \frac{\alpha }{2}\right) \left\vert r\right\rangle +\sin \left( 
\frac{\alpha }{2}\right) \left\vert a\right\rangle \text{,}
\end{equation}%
where the angle $\alpha $ characterizes the overlap between $\left\vert \bar{%
q}\right\rangle $ and the searched state $\left\vert a\right\rangle $,%
\begin{equation}
\sin \left( \frac{\alpha }{2}\right) \overset{\text{def}}{=}\frac{1}{\sqrt{N}%
}\equiv \left\langle a|\bar{q}\right\rangle \text{,}  \label{sin}
\end{equation}%
while the state $\left\vert r\right\rangle $ is defined as,%
\begin{equation}
\left\vert r\right\rangle \overset{\text{def}}{=}\sqrt{\frac{1}{N-1}}%
\sum_{s\neq a}\left\vert s\right\rangle \text{.}
\end{equation}%
Given the initial input, a single-step of Grover's algorithm is
characterized by a rotation by $\alpha $ in the two-dimensional space
spanned by $\left\vert r\right\rangle $ and $\left\vert a\right\rangle $.
The effect of the algorithm after $m$-steps leads to the state $\left\vert
\psi _{\text{G}}\left( m\right) \right\rangle $,%
\begin{equation}
\left\vert \psi _{\text{G}}\left( m\right) \right\rangle =\cos \left[ \left(
m+\frac{1}{2}\right) \alpha \right] \left\vert r\right\rangle +\sin \left[
\left( m+\frac{1}{2}\right) \alpha \right] \left\vert a\right\rangle \text{.}
\label{output}
\end{equation}%
In the limit of $N\gg 1$, the number of steps $\bar{m}$ for which $%
\left\vert \psi _{\text{G}}\left( \bar{m}\right) \right\rangle $ coincides
with the searched state $\left\vert a\right\rangle $ (with success
probability equal to $1$) is approximately given by,%
\begin{equation}
\bar{m}\overset{N\gg 1}{\simeq }\frac{\pi }{4}\sqrt{N}\text{.}
\label{finalino}
\end{equation}%
Equation (\ref{finalino}) can be obtained by imposing that $\left( \bar{m}+%
\frac{1}{2}\right) \alpha =\frac{\pi }{2}$ and by observing that when $N\gg
1 $, Eq. (\ref{sin}) implies that $\alpha \simeq \frac{2}{\sqrt{N}}$.

Although Grover's algorithm evolves with discrete $m$, in the limit of $N\gg
1$ the output state (\ref{output}) can be approximated by a quantum
wave-state $\left\vert \psi \left( \theta \right) \right\rangle $ depending
on a continuous parameter $\theta $. Indeed, considering the following
formal substitutions,%
\begin{equation}
\left( m+\frac{1}{2}\right) \alpha \rightarrow \theta \text{, }\left\vert
a\right\rangle \rightarrow \left\vert 0\right\rangle \text{, }\left\vert
r\right\rangle \rightarrow \frac{1}{\sqrt{N-1}}\sum_{k=1}^{N-1}\left\vert
k\right\rangle \text{,}
\end{equation}%
the state $\left\vert \psi \left( \theta \right) \right\rangle $ reads,%
\begin{equation}
\left\vert \psi \left( \theta \right) \right\rangle =\sum_{k=0}^{N-1}\sqrt{%
p_{k}\left( \theta \right) }\left\vert k\right\rangle \text{,}  \label{conti}
\end{equation}%
where,%
\begin{equation}
\left\langle k|k^{\prime }\right\rangle =\delta _{kk^{\prime }}\text{, }%
p_{0}\left( \theta \right) \overset{\text{def}}{=}\sin ^{2}\theta \text{
and, }p_{j}\left( \theta \right) \overset{\text{def}}{=}\frac{\cos
^{2}\theta }{N-1}\text{ with }j\neq 0\text{.}  \label{probabilities}
\end{equation}%
The $N$-dimensional probability distribution vector $\vec{p}\equiv \left(
p_{0}\left( \theta \right) \text{, }p_{1}\left( \theta \right) \text{,..., }%
p_{N-1}\left( \theta \right) \right) $ with $p_{k}\left( \theta \right) $
defined in (\ref{probabilities}) can be regarded as a path characterizing
Grover's algorithm on a suitable probability space. In what follows, we show
that such a path is indeed a geodesic path for which the quantum Fisher
information action functional achieves an extremal value.

\section{Information Geometry and Grover's Algorithm}

In this section, we show that Grover's dynamics leading to (\ref{conti})
corresponds to the shortest path from an information geometric point of view.

\subsection{The Parametric Pure State Model}

Consider the parametric density operator $\rho _{\theta }$ constructed using
the quantum state $\left\vert \psi \left( \theta \right) \right\rangle
\equiv \left\vert \psi _{\theta }\right\rangle $ in (\ref{conti}),%
\begin{equation}
\rho _{\theta }\overset{\text{def}}{=}\left\vert \psi _{\theta
}\right\rangle \left\langle \psi _{\theta }\right\vert =\sum_{k\text{, }%
j=0}^{N-1}\left[ p_{k}\left( \theta \right) p_{j}\left( \theta \right) %
\right] ^{\frac{1}{2}}e^{-i\left[ \phi _{j}\left( \theta \right) -\phi
_{k}\left( \theta \right) \right] }\left\vert k\right\rangle \left\langle
j\right\vert \text{,}  \label{do}
\end{equation}%
where the pure state $\left\vert \psi _{\theta }\right\rangle $ is
normalized to one. Since $\rho _{\theta }$ is a pure state, it satisfies the
following relations,%
\begin{equation}
\rho _{\theta }^{2}=\rho _{\theta }\text{ and, }\rho _{\theta }=\sqrt{\rho
_{\theta }}\text{.}
\end{equation}%
We assume that the notion of distinguishability on the manifold of density
operators defined in (\ref{do}) is provided by the Wigner-Yanase quantum
Fisher metric defined as,%
\begin{equation}
\left[ I_{WY}^{\left( \text{quantum}\right) }\right] _{ij}\left( \rho
_{\theta }\right) \overset{\text{def}}{=}4\text{Tr}\left[ \left( \partial
_{i}\sqrt{\rho _{\theta }}\right) \left( \partial _{j}\sqrt{\rho _{\theta }}%
\right) \right] =4\text{Tr}\left[ \left( \partial _{i}\rho _{\theta }\right)
\left( \partial _{j}\rho _{\theta }\right) \right] \text{,}  \label{def}
\end{equation}%
where in the multi-parametric case $\theta \equiv \left( \theta ^{1}\text{%
,..., }\theta ^{l}\right) $ and $\partial _{i}\overset{\text{def}}{=}\frac{%
\partial }{\partial \theta ^{i}}$ with $i=1$,..., $l$. The second equality
in (\ref{def}) holds because $\rho _{\theta }$ is a pure state in the case
being considered. After some algebraic manipulations and observing that the
normalization condition for $\left\vert \psi _{\theta }\right\rangle $
implies that,%
\begin{equation}
\left\langle \partial _{j}\psi _{\theta }|\psi _{\theta }\right\rangle
=-\left\langle \psi _{\theta }|\partial _{j}\psi _{\theta }\right\rangle 
\text{,}
\end{equation}%
the Wigner-Yanase quantum Fisher metric becomes,%
\begin{equation}
\left[ I_{WY}^{\left( \text{quantum}\right) }\right] _{ij}\left( \rho
_{\theta }\right) =4\left[ \left\langle \partial _{i}\psi _{\theta
}|\partial _{j}\psi _{\theta }\right\rangle +\left\langle \partial _{i}\psi
_{\theta }|\psi _{\theta }\right\rangle \left\langle \partial _{j}\psi
_{\theta }|\psi _{\theta }\right\rangle \right] \text{.}
\end{equation}%
Thus, the infinitesimal line element on the manifold of density operators in
(\ref{do}) reads,%
\begin{equation}
ds_{\text{WY}}^{2}=\left[ I_{WY}^{\left( \text{quantum}\right) }\right]
_{ij}\left( \rho _{\theta }\right) d\theta _{i}d\theta _{j}=4\left[
\left\langle \partial _{i}\psi _{\theta }|\partial _{j}\psi _{\theta
}\right\rangle +\left\langle \partial _{i}\psi _{\theta }|\psi _{\theta
}\right\rangle \left\langle \partial _{j}\psi _{\theta }|\psi _{\theta
}\right\rangle \right] d\theta _{i}d\theta _{j}\text{.}  \label{line}
\end{equation}%
Defining $\gamma _{ij}$ and $\sigma _{ij}$ as,%
\begin{equation}
\gamma _{ij}\overset{\text{def}}{=}\func{Re}\left[ \left\langle \partial
_{i}\psi _{\theta }|\partial _{j}\psi _{\theta }\right\rangle \right] \text{
and }\sigma _{ij}\overset{\text{def}}{=}\func{Im}\left[ \left\langle
\partial _{i}\psi _{\theta }|\partial _{j}\psi _{\theta }\right\rangle %
\right] \text{, }  \label{z-1}
\end{equation}%
respectively, it follows that%
\begin{equation}
\gamma _{ij}=\gamma _{ji}\text{ and, }\sigma _{ij}=-\sigma _{ji}\text{. }
\end{equation}%
Because of the asymmetry of $\sigma _{ij}$, we have%
\begin{equation}
\sigma _{ij}d\theta _{i}d\theta _{j}=-\sigma _{ji}d\theta _{i}d\theta
_{j}\equiv -\sigma _{ji}d\theta _{j}d\theta _{i}=-\sigma _{ij}d\theta
_{i}d\theta _{j}\text{,}
\end{equation}%
that is,%
\begin{equation}
\sigma _{ij}d\theta _{i}d\theta _{j}=0\text{.}  \label{za}
\end{equation}%
Therefore, it turns out that $ds_{\text{WY}}^{2}$ in (\ref{line}) becomes%
\begin{equation}
ds_{\text{WY}}^{2}=\left[ I_{WY}^{\left( \text{quantum}\right) }\right]
_{ij}\left( \rho _{\theta }\right) d\theta _{i}d\theta _{j}=4\left[ \func{Re}%
\left\langle \partial _{i}\psi _{\theta }|\partial _{j}\psi _{\theta
}\right\rangle +\left\langle \partial _{i}\psi _{\theta }|\psi _{\theta
}\right\rangle \left\langle \partial _{j}\psi _{\theta }|\psi _{\theta
}\right\rangle \right] d\theta _{i}d\theta _{j}\text{.}  \label{line2}
\end{equation}%
We remark that $ds_{\text{WY}}^{2}$ in (\ref{line2}) is exactly four times
the Fubini-Study infinitesimal line element is given by \cite{provost},%
\begin{equation}
ds_{\text{FS}}^{2}=g_{ij}^{\left( \text{FS}\right) }\left( \theta \right)
d\theta _{i}d\theta _{j}=\left\Vert d\psi \right\Vert ^{2}-\left\vert
\left\langle \psi |d\psi \right\rangle \right\vert ^{2}=1-\left\vert
\left\langle \psi ^{\prime }|\psi \right\rangle \right\vert ^{2}\text{,}
\label{ray}
\end{equation}%
where $\left\vert \psi \right\rangle =\left\vert \psi \left( \theta \right)
\right\rangle \equiv \left\vert \psi _{\theta }\right\rangle $, $\left\vert
\psi ^{\prime }\right\rangle =\left\vert \psi \left( \theta +d\theta \right)
\right\rangle \equiv \left\vert \psi _{\theta +d\theta }\right\rangle $ and
the Fubini-Study metric $g_{ij}^{\left( \text{FS}\right) }\left( \theta
\right) $ reads,%
\begin{equation}
g_{ij}^{\left( \text{FS}\right) }\left( \theta \right) \equiv g_{ij}\left(
\theta \right) \overset{\text{def}}{=}\left[ \func{Re}\left\langle \partial
_{i}\psi _{\theta }|\partial _{j}\psi _{\theta }\right\rangle +\left\langle
\partial _{i}\psi _{\theta }|\psi _{\theta }\right\rangle \left\langle
\partial _{j}\psi _{\theta }|\psi _{\theta }\right\rangle \right] \text{.}
\end{equation}%
The Fubini-Study metric is a gauge invariant metric on the manifold of
Hilbert-space rays (projective Hilbert space). Our analysis explicitly
recognizes that the Fubini-Study metric is a quantum version of the Fisher
metric \cite{hiroshi}.

\subsection{Grover's Information Geometric Dynamics}

Substituting (\ref{ray}) into (\ref{line2}) and using (\ref{conti}) together
with the normalization condition on the probabilities in (\ref{probabilities}%
), after some straightforward algebra it turns out that the infinitesimal
Wigner-Yanase line element reads (for further details, see Appendix A),%
\begin{equation}
ds_{\text{WY}}^{2}=\left\{ \sum_{k=0}^{N-1}\frac{\dot{p}_{k}^{2}}{p_{k}}+4%
\left[ \sum_{k=0}^{N-1}p_{k}\dot{\phi}_{k}^{2}-\left( \sum_{k=0}^{N-1}p_{k}%
\dot{\phi}_{k}\right) ^{2}\right] \right\} d\theta ^{2}\text{,}
\end{equation}%
where,%
\begin{equation}
\dot{p}_{k}\overset{\text{def}}{=}\frac{dp_{k}\left( \theta \right) }{%
d\theta }\text{ and, }\dot{\phi}_{k}\overset{\text{def}}{=}\frac{d\phi
_{k}\left( \theta \right) }{d\theta }\text{.}
\end{equation}%
Recall that Grover's algorithm consists only of a sequence of unitary
transformations on a pure state. In the continuous approximation, such
sequence of quantum unitary operations leads to the output state $\left\vert
\psi \left( \theta \right) \right\rangle $ in (\ref{conti}). Such quantum
mechanical wave-vector is such that its Fisher information function $%
\mathcal{F}\left( \theta \right) $ \cite{brandt2} is independent of the
parameter\textbf{\ }$\theta $\textbf{\ }(it is a constant quantity),%
\begin{equation}
\mathcal{F}\left( \theta \right) \overset{\text{def}}{=}%
\sum_{k=0}^{N-1}p_{k}\left( \frac{\partial \log p_{k}}{\partial \theta }%
\right) ^{2}=\sum_{k=0}^{N-1}\frac{\dot{p}_{k}^{2}}{p_{k}}%
=4\sum_{k=0}^{N-1}\left( \frac{\partial \sqrt{p_{k}}}{\partial \theta }%
\right) ^{2}=4\text{. }  \label{FF}
\end{equation}%
We point out that the general expression of the Fisher information function%
\textbf{\ }$\mathcal{F}\left( \theta \right) $\textbf{\ }in (\ref{FF}) is
invariant under unitary transformations. This can be explained as follows.
Consider a normalized pure state\textbf{\ }$\left\vert \psi \left( \theta
\right) \right\rangle $\textbf{\ }given by,%
\begin{equation}
\left\vert \psi \left( \theta \right) \right\rangle \overset{\text{def}}{=}%
\sum_{k=0}^{N-1}\sqrt{p_{k}\left( \theta \right) }e^{i\phi _{k}\left( \theta
\right) }\left\vert k\right\rangle \text{.}  \label{fifi}
\end{equation}%
Eq. (\ref{fifi})\ implies that,%
\begin{equation}
\left\vert \psi _{\theta }\left( m\right) \right\vert =\sqrt{p_{m}\left(
\theta \right) }\text{,}
\end{equation}%
where\textbf{\ }$\psi _{\theta }\left( m\right) \overset{\text{def}}{=}%
\left\langle m|\psi \left( \theta \right) \right\rangle $ with $\left\langle
m|k\right\rangle =\delta _{mk}$\textbf{\ }and the Fisher information
function reads,%
\begin{equation}
\mathcal{F}\left( \theta \right) =4\sum_{m=0}^{N-1}\left( \frac{\partial
\left\vert \psi _{\theta }\left( m\right) \right\vert }{\partial \theta }%
\right) ^{2}=\sum_{m=0}^{N-1}\frac{\dot{p}_{m}^{2}}{p_{m}}\text{.}
\label{fifi1}
\end{equation}%
\textbf{\ }Under unitary transformations\textbf{\ }$U$\textbf{\ }mapping%
\textbf{\ }$\left\vert \psi \left( \theta \right) \right\rangle $\ to\textbf{%
\ }$\left\vert \psi ^{\prime }\left( \theta \right) \right\rangle
=U\left\vert \psi \left( \theta \right) \right\rangle $\textbf{, }the
transformed Fisher information function\textbf{\ }$\mathcal{F}^{\prime
}\left( \theta \right) $\textbf{\ }becomes,%
\begin{equation}
\mathcal{F}^{\prime }\left( \theta \right) =4\sum_{m=0}^{N-1}\left( \frac{%
\partial \left\vert \psi _{\theta }^{\prime }\left( u_{m}\right) \right\vert 
}{\partial \theta }\right) ^{2}=\sum_{m=0}^{N-1}\frac{\dot{p}_{m}^{2}}{p_{m}}%
\text{,}  \label{fifi2}
\end{equation}%
where\textbf{\ }$\psi _{\theta }^{\prime }\left( u_{m}\right) \overset{\text{%
def}}{=}\left\langle u_{m}|\psi ^{\prime }\left( \theta \right)
\right\rangle $\textbf{\ }with\textbf{\ }$\left\langle
u_{m}|u_{k}\right\rangle =\delta _{mk}$\textbf{\ }and\textbf{\ }$\left\vert
u_{k}\right\rangle \overset{\text{def}}{=}U\left\vert k\right\rangle $%
\textbf{.} Thus, from (\ref{fifi1}) and (\ref{fifi2}) it turns out that the
Fisher information function remains unchanged under unitary operations. In
the rest of the manuscript, we refer to Eq. (\ref{FF}) as the "\emph{%
parametric-independence constraint}" on the Fisher information function.

As a side remark, we emphasize that within a statistical inference viewpoint
extended to the quantum framework, the statistical notion of Fisher
information function resembles a generalized mechanical notion of kinetic
energy with respect to a given statistical parameter ($\theta $, in our
case) regarded as temporal or spatial shift. Specifically, it can be shown
that the following relation holds \cite{luo33},%
\begin{equation}
\mathcal{K}\left( \theta \right) =\frac{1}{4}\mathcal{F}\left( \theta
\right) +\sum_{k=0}^{N-1}J_{\theta }^{2}\left( k\right) \left\vert \psi
_{\theta }\left( k\right) \right\vert ^{2}\text{,}  \label{relation1}
\end{equation}%
with\textbf{\ }$\mathcal{F}\left( \theta \right) $\textbf{\ }defined as in (%
\ref{fifi1}) and\textbf{\ }$\psi _{\theta }\left( k\right) =\left\langle
k|\psi _{\theta }\right\rangle $ and $p_{k}\left( \theta \right) =\left\vert
\psi _{\theta }\left( k\right) \right\vert ^{2}$. The quantity $\mathcal{K}%
\left( \theta \right) $ denotes the kinetic energy of the wavefunction with
respect to the parameter $\theta $ while $J_{\theta }\left( k\right) $ is a
statistical analogue of the normalized quantum mechanical current density
with respect to $\theta $. They are defined as \cite{luo33},%
\begin{equation}
\mathcal{K}\left( \theta \right) \overset{\text{def}}{=}\sum_{k=0}^{N-1}%
\left\vert \frac{\partial \psi _{\theta }\left( k\right) }{\partial \theta }%
\right\vert ^{2}\text{,}
\end{equation}%
and,%
\begin{equation}
J_{\theta }\left( k\right) \overset{\text{def}}{=}\frac{1}{2i\left\vert \psi
_{\theta }\left( k\right) \right\vert ^{2}}\left( \frac{\partial \psi
_{\theta }\left( k\right) }{\partial \theta }\psi _{\theta }^{\ast }\left(
k\right) -\psi _{\theta }\left( k\right) \frac{\partial \psi _{\theta
}^{\ast }\left( k\right) }{\partial \theta }\right) \text{,}
\end{equation}%
respectively. The symbol "$\ast $" denotes complex conjugation.\textbf{\ }For%
\textbf{\ }$\psi _{\theta }\left( k\right) =\sqrt{p_{k}\left( \theta \right) 
}e^{i\phi _{k}\left( \theta \right) }$\textbf{,} we obtain that%
\begin{equation}
\mathcal{K}\left( \theta \right) =\left\langle \dot{\psi}_{\theta }|\dot{\psi%
}_{\theta }\right\rangle \text{, }\mathcal{F}\left( \theta \right)
=\sum_{k=0}^{N-1}\frac{\dot{p}_{k}^{2}}{p_{k}}\text{ and, }J_{\theta }\left(
k\right) =\dot{\phi}_{k}\left( \theta \right) \text{,}
\end{equation}%
\textbf{\ }with\textbf{\ }$\dot{\psi}_{\theta }\overset{\text{def}}{=}\frac{%
d\psi \left( \theta \right) }{d\theta }$. Therefore, Eq. (\ref{relation1})
reads%
\begin{equation}
\left\langle \dot{\psi}_{\theta }|\dot{\psi}_{\theta }\right\rangle =\frac{1%
}{4}\sum_{k=0}^{N-1}\frac{\dot{p}_{k}^{2}}{p_{k}}+\sum_{k=0}^{N-1}p_{k}\dot{%
\phi}_{k}^{2}\text{.}  \label{relation2}
\end{equation}%
Since in Grover's case\textbf{\ }$\mathcal{F}\left( \theta \right) =4$%
\textbf{\ }and\textbf{\ }$J_{\theta }\left( k\right) =0$\textbf{\ }for any%
\textbf{\ }$k=0$\textbf{,..., }$N-1$\textbf{, }condition (\ref{relation2})
becomes%
\begin{equation}
\mathcal{K}\left( \theta \right) =\left\langle \dot{\psi}_{\theta }|\dot{\psi%
}_{\theta }\right\rangle =1\text{.}
\end{equation}%
We conclude that from a statistical quantum inference viewpoint, Grover's
algorithm is characterized by a constant statistical kinetic energy where no
"statistical dissipation" occurs.

The geodesic path related to Grover's algorithm is found by minimizing the
action $\mathcal{S}\left[ p_{k}\left( \theta \right) \right] $ defined as,%
\begin{equation}
\mathcal{S}\left[ p_{k}\left( \theta \right) \right] \overset{\text{def}}{=}%
\int \sqrt{ds_{\text{WY}}^{2}}=\int \mathcal{L}\left( \dot{p}_{k}\left(
\theta \right) \text{, }p_{k}\left( \theta \right) \right) d\theta \text{,}
\end{equation}%
where the Lagrangian-like quantity $\mathcal{L}\left( \dot{p}_{k}\left(
\theta \right) \text{, }p_{k}\left( \theta \right) \right) $ reads,%
\begin{equation}
\mathcal{L}\left( \dot{p}_{k}\left( \theta \right) \text{, }p_{k}\left(
\theta \right) \right) =\left[ \sum_{k=0}^{N-1}\frac{\dot{p}_{k}^{2}\left(
\theta \right) }{p_{k}\left( \theta \right) }\right] ^{\frac{1}{2}}\text{,}
\label{lagrangian}
\end{equation}%
given the normalization constraint on the parametric probabilities $%
p_{k}\left( \theta \right) $,%
\begin{equation}
\sum_{k=0}^{N-1}p_{k}\left( \theta \right) =1\text{.}  \label{nc}
\end{equation}%
For the sake of simplicity, consider the change of variable $p_{k}\left(
\theta \right) \rightarrow q_{k}^{2}\left( \theta \right) $ \cite{wootters}.
Then, using the Lagrange multipliers method, the new action $\mathcal{S}%
^{\prime }\left[ q_{k}\left( \theta \right) \right] $ to minimize becomes ,%
\begin{equation}
\mathcal{S}^{\prime }\left[ q_{k}\left( \theta \right) \right] =\int 
\mathcal{L}^{\prime }\left( \dot{q}_{k}\left( \theta \right) \text{, }%
q_{k}\left( \theta \right) \right) d\theta =\int \left\{ \left[
4\sum_{k=1}^{N}\dot{q}_{k}^{2}\left( \theta \right) \right] ^{\frac{1}{2}%
}-\lambda \left( \sum_{k=1}^{N}q_{k}^{2}\left( \theta \right) -1\right)
\right\} d\theta \text{,}
\end{equation}%
where $\lambda $ is the Lagrange multiplier and the new Lagrangian-like
quantity is given by,%
\begin{equation}
\mathcal{L}^{\prime }\left( \dot{q}_{k}\left( \theta \right) \text{, }%
q_{k}\left( \theta \right) \right) =\left[ 4\sum_{k=1}^{N}\dot{q}%
_{k}^{2}\left( \theta \right) \right] ^{\frac{1}{2}}-\lambda \left(
\sum_{k=1}^{N}q_{k}^{2}\left( \theta \right) -1\right) \text{.}  \label{A}
\end{equation}%
The path minimizing the action $\mathcal{S}^{\prime }\left[ q_{k}\left(
\theta \right) \right] $ satisfies the "\emph{actuality constraint}",%
\begin{equation}
\frac{\delta \mathcal{S}^{\prime }\left[ q_{k}\left( \theta \right) \right] 
}{\delta q_{k}\left( \theta \right) }=0\text{,}  \label{minima}
\end{equation}%
leading to the Euler-Lagrange (EL) equation,%
\begin{equation}
\frac{d}{d\theta }\left( \frac{\partial \mathcal{L}^{\prime }\left( \dot{q}%
_{k}\left( \theta \right) \text{, }q_{k}\left( \theta \right) \right) }{%
\partial \dot{q}_{k}}\right) -\frac{\partial \mathcal{L}^{\prime }\left( 
\dot{q}_{k}\left( \theta \right) \text{, }q_{k}\left( \theta \right) \right) 
}{\partial q_{k}}=0\text{.}  \label{B}
\end{equation}%
From (\ref{A}), it follows that%
\begin{equation}
\frac{d}{d\theta }\left( \frac{\partial \mathcal{L}^{\prime }}{\partial \dot{%
q}_{k}}\right) =\frac{4\ddot{q}_{k}}{\left( 4\overset{N}{\underset{k=1}{\sum 
}}\dot{q}_{k}^{2}\right) ^{\frac{1}{2}}}-\frac{\dot{q}_{k}\ddot{q}_{k}}{%
\left( 4\overset{N}{\underset{k=1}{\sum }}\dot{q}_{k}^{2}\right) ^{\frac{1}{2%
}}}\frac{1}{\left( 4\overset{N}{\underset{k=1}{\sum }}\dot{q}_{k}^{2}\right) 
}4\dot{q}_{k}\text{,}  \label{C}
\end{equation}%
and,%
\begin{equation}
\frac{\partial \mathcal{L}^{\prime }}{\partial q_{k}}=-2\lambda q_{k}\text{.}
\label{D}
\end{equation}%
Thus, substituting (\ref{D}) and (\ref{C}) into (\ref{B}), the EL\ equation
becomes%
\begin{equation}
\frac{d^{2}q_{k}\left( \theta \right) }{d\theta ^{2}}-\frac{\mathcal{\dot{L}}%
\left( \dot{q}_{k}\left( \theta \right) \text{, }q_{k}\left( \theta \right)
\right) }{\mathcal{L}\left( \dot{q}_{k}\left( \theta \right) \text{, }%
q_{k}\left( \theta \right) \right) }\frac{dq_{k}\left( \theta \right) }{%
d\theta }+\frac{\lambda }{2}\mathcal{L}\left( \dot{q}_{k}\left( \theta
\right) \text{, }q_{k}\left( \theta \right) \right) q_{k}\left( \theta
\right) =0\text{,}  \label{EL}
\end{equation}%
where $\mathcal{\dot{L}}\left( \dot{q}_{k}\left( \theta \right) \text{, }%
q_{k}\left( \theta \right) \right) =\frac{d\mathcal{L}\left( \dot{q}%
_{k}\left( \theta \right) \text{, }q_{k}\left( \theta \right) \right) }{%
d\theta }$ with $\mathcal{L}\left( \dot{q}_{k}\left( \theta \right) \text{, }%
q_{k}\left( \theta \right) \right) $ given in (\ref{lagrangian}) with $%
q_{k}^{2}\left( \theta \right) =p_{k}\left( \theta \right) $. Using (\ref{FF}%
), it follows that $\mathcal{L}\left( \dot{q}_{k}\left( \theta \right) \text{%
, }q_{k}\left( \theta \right) \right) $ is constant and equals two while $%
\mathcal{\dot{L}}\left( \dot{q}_{k}\left( \theta \right) \text{, }%
q_{k}\left( \theta \right) \right) $ equals zero. It finally follows that%
\begin{equation}
q_{0}\left( \theta \right) =\sin \theta \text{ and, }q_{\bar{k}}\left(
\theta \right) =\frac{\cos \theta }{\sqrt{N-1}}\text{,}
\end{equation}%
where $\bar{k}=1$,..., $N-1$ fulfil the Eq. (\ref{EL}) with $\lambda $ equal
to one in order to satisfy the normalization condition (\ref{nc}). We then
conclude that the $N$-dimensional probability vector $\vec{p}\equiv \left(
p_{0}\left( \theta \right) \text{, }p_{1}\left( \theta \right) \text{,..., }%
p_{N-1}\left( \theta \right) \right) $ with $p_{k}\left( \theta \right) $
defined in (\ref{probabilities}) is a geodesic path for which the quantum
Fisher information action functional $\mathcal{S}^{\prime }\left[
q_{k}\left( \theta \right) \right] $ achieves an extremal value. We also
stress that such a path satisfies both the parametric-independence and
actuality constraints in\ Eqs. (\ref{FF}) and (\ref{minima}), respectively.

\section{Deviations from Grover's Algorithm}

As pointed out earlier, Grover's search algorithm relies on two
key-assumptions: 1) Grover's iteration must be characterized only by means
of unitary transformations acting on a uniform amplitude input state. This
leads to the parametric-independence constraint on the Fisher information
function (see Eq. (\ref{FF})); 2) the expressions for the components of the
probability distribution vectors associated to Grover's algorithm are such
that the quantum Fisher information action functional is minimized when
evaluated on such paths (see Eq. (\ref{minima})). In summary, Grover's paths
are \emph{actual} \emph{unitary} \emph{probability paths of constant Fisher
information}. When one of these two requirements (parametric-invariance and
actuality) do not hold anymore, we are led to "deviations"\ from Grover's
algorithm which should be somehow less efficient than Grover's one.

In what follows, we discuss unitary deviations from Grover's algorithm and
attempt to quantify, whenever possible, different aspects from an
information geometric point of view on manifolds where the statistical
parameter $\theta $ can be regarded as a local coordinate parameterizing the
density operator $\rho _{\theta }$.

\subsection{Grover's Model: Actuality and Parametric-Independence}

Recall that the probability distribution vectors associated with Grover's
algorithm is given by,%
\begin{equation}
\vec{p}_{\text{G}}\equiv \left( \sin ^{2}\theta \text{, }\frac{\cos
^{2}\theta }{N-1}\text{,..., }\frac{\cos ^{2}\theta }{N-1}\right) \text{.}
\label{p}
\end{equation}%
Eq. (\ref{p}) leads to the constancy of the Fisher information function $%
\mathcal{F}\left( \theta \right) $,%
\begin{equation}
\mathcal{F}\left( \theta \right) =\sum_{k=0}^{N-1}\frac{1}{p_{k}}\left( 
\frac{dp_{k}\left( \theta \right) }{d\theta }\right) ^{2}=4\text{.}
\end{equation}%
The geodesic equation in (\ref{GE}) for Grover's statistical model reads,%
\begin{equation}
\ddot{\theta}+\frac{1}{2\mathcal{F}\left( \theta \right) }\frac{d\mathcal{F}%
\left( \theta \right) }{d\theta }\dot{\theta}^{2}=0\text{,}  \label{lde}
\end{equation}%
with $\dot{\theta}\overset{\text{def}}{=}\frac{d\theta \left( \tau \right) }{%
d\tau }$. Since $\mathcal{F}\left( \theta \right) $ is constant, the
geodesic equation reads $\ddot{\theta}=0$. Assuming that the boundary
conditions are given by $\theta \left( \tau _{i}\right) =\theta _{i}$ and $%
\theta \left( \tau _{f}\right) =\theta _{f}$, integration of (\ref{lde})
leads to the geodesic path $\theta _{\text{G}}\left( \tau \right) $,%
\begin{equation}
\theta _{\text{G}}\left( \tau \right) =\frac{\theta _{i}\tau _{f}-\theta
_{f}\tau _{i}}{\tau _{f}-\tau _{i}}+\left( \frac{\theta _{f}-\theta _{i}}{%
\tau _{f}-\tau _{i}}\right) \tau \text{.}
\end{equation}%
The geodesic path $\theta _{\text{G}}\left( \tau \right) $ can be regarded
as a \emph{continuous} succession of intermediate states connecting $\theta
_{i}$ to $\theta _{f}$. The geodesic motion occurs with speed $v_{\text{G}%
}\left( \tau \right) $,%
\begin{equation}
v_{\text{Grover}}\left( \tau \right) \overset{\text{def}}{=}\frac{d\theta _{%
\text{Grover}}\left( \tau \right) }{d\tau }\text{,}
\end{equation}%
satisfying the normalization relation,%
\begin{equation}
\left[ g_{lm}\left( \theta \right) \frac{d\theta ^{l}\left( \tau \right) }{%
d\tau }\frac{d\theta ^{m}\left( \tau \right) }{d\tau }\right] ^{\frac{1}{2}%
}=1\text{,}  \label{nc0}
\end{equation}%
with $l=m=1$ and $g_{11}\left( \theta \right) =4$. Equation (\ref{nc0})
implies that the temporal duration $\Delta \tau $ needed to navigate the
geodesic path connecting $\theta _{i}$ to $\theta _{f}$ is given by,%
\begin{equation}
\Delta \tau \overset{\text{def}}{=}\tau _{f}-\tau _{i}=2\left( \theta
_{f}-\theta _{i}\right) \text{.}
\end{equation}%
We point out that $\Delta \tau $ can also be viewed as the length $L_{\text{G%
}}\left( \theta _{i}\text{, }\theta _{f}\right) $ of the geodesic path
connecting $\theta _{i}$ to $\theta _{f}$,%
\begin{equation}
L_{\text{G}}\left( \theta _{i}\text{, }\theta _{f}\right) \overset{\text{def}%
}{=}\int_{\theta _{i}}^{\theta _{f}}\left[ g_{lm}\left( \theta \right)
d\theta ^{l}d\theta ^{m}\right] ^{\frac{1}{2}}=\int_{\tau _{i}}^{\tau
_{f}^{\prime }}d\tau \left[ g_{lm}\left( \theta \right) \frac{d\theta
^{l}\left( \tau \right) }{d\tau }\frac{d\theta ^{m}\left( \tau \right) }{%
d\tau }\right] ^{\frac{1}{2}}\text{.}
\end{equation}%
Indeed, in the case being considered $L_{\text{G}}\left( \theta _{i}\text{, }%
\theta _{f}\right) $\textbf{\ }reads%
\begin{equation}
L_{\text{G}}\left( \theta _{i}\text{, }\theta _{f}\right) \overset{\text{def}%
}{=}\int_{\theta _{i}}^{\theta _{f}}\sqrt{4}d\theta =2\left( \theta
_{f}-\theta _{i}\right) \equiv \Delta \tau \text{.}
\end{equation}

\subsection{Deviation from Grover's Model: Neither Actuality Nor
Parametric-Independence}

In this example\textbf{,} we consider a properly normalized probability
distribution vector given by,%
\begin{equation}
\vec{p}_{\text{Model-II}}\equiv \left( \theta ^{2}\text{, }\frac{1-\theta
^{2}}{N-1}\text{,..., }\frac{1-\theta ^{2}}{N-1}\right) \text{.}
\end{equation}%
Notice that when $\theta \ll 1$, $\vec{p}_{\text{Grover}}\approx \vec{p}_{%
\text{Model-II}}$. The Fisher information function $\mathcal{F}\left( \theta
\right) $ for the modified Grover's statistical model is not constant (no
parametric-independence) and reads,%
\begin{equation}
\mathcal{F}\left( \theta \right) =\sum_{k=0}^{N-1}\frac{1}{p_{k}}\left( 
\frac{dp_{k}\left( \theta \right) }{d\theta }\right) ^{2}=\frac{4}{1-\theta
^{2}}\text{.}  \label{ft}
\end{equation}%
We also remark that $\vec{p}_{\text{Model-II}}$ with $\mathcal{F}\left(
\theta \right) $ in (\ref{ft}) does not satisfy the EL\ equation in (\ref{EL}%
). According to the principle of least action, this means that $\vec{p}_{%
\text{Model-II}}$ can be regarded as a possible but not an actual (not
extremal) probability path. Furthermore, the geodesic equation (\ref{GE})
for the modified Grover's statistical model becomes,%
\begin{equation}
\ddot{\theta}+\frac{\theta }{1-\theta ^{2}}\dot{\theta}^{2}=0\text{,}
\label{node}
\end{equation}%
where $\dot{\theta}=\frac{d\theta \left( \tau \right) }{d\tau }$. Eq. (\ref%
{node}) is a \emph{nonlinear} second order ordinary differential equation.
Setting $\dot{\theta}=\frac{d\theta \left( \tau \right) }{d\tau }=y\left(
\theta \right) $, Eq. (\ref{node}) becomes a first order differential
equation,%
\begin{equation}
yy^{\prime }+\frac{\theta }{1-\theta ^{2}}y^{2}=0\text{, }  \label{de2}
\end{equation}%
where $y^{\prime }=\frac{dy\left( \theta \right) }{d\theta }$. Assuming $%
y\neq 0$, integration of (\ref{de2}) implies that%
\begin{equation}
y\left( \theta \right) =C\left( 1-\theta ^{2}\right) ^{\frac{1}{2}}\text{,}
\end{equation}%
where $C\in 
\mathbb{R}
$ is an integration constant. Since $y\left( \theta \right) =\frac{d\theta
\left( \tau \right) }{d\tau }$, it finally turns out that the general
solution of Eq. (\ref{node}) is given by,%
\begin{equation}
\theta \left( \tau \right) =\sin \left[ C_{0}\left( \tau +C_{1}\right) %
\right] \text{,}
\end{equation}%
where $C_{0}$ and $C_{1}$ are \emph{real} integration constants whose
explicit expressions can be obtained once the boundary conditions are fixed.
Assuming $\theta \left( \tau _{i}\right) =\theta _{i}$ and $\theta \left(
\tau _{f}\right) =\theta _{f}$, the expression for the geodesic path $\theta
_{\text{Model-II}}\left( \tau \right) $ reads,%
\begin{equation}
\theta _{\text{Model-II}}\left( \tau \right) =\sin \left[ \left( \frac{%
\arcsin \theta _{f}-\arcsin \theta _{i}}{\tau _{f}^{\prime }-\tau
_{i}^{\prime }}\right) \left( \frac{\tau _{f}^{\prime }\arcsin \theta
_{i}-\tau _{i}^{\prime }\arcsin \theta _{f}}{\arcsin \theta _{f}-\arcsin
\theta _{i}}+\tau \right) \right] \text{.}
\end{equation}%
The geodesic path $\theta \left( \tau \right) $ can be regarded as a \emph{%
continuous} succession of intermediate states connecting $\theta _{i}$ to $%
\theta _{f}$. The geodesic motion occurs with speed $v_{\text{Model-II}%
}\left( \tau \right) $,%
\begin{equation}
v_{\text{Model-II}}\left( \tau \right) \overset{\text{def}}{=}\frac{d\theta
_{\text{Model-II}}\left( \tau \right) }{d\tau }\text{,}
\end{equation}%
satisfying the normalization relation,%
\begin{equation}
\left[ g_{lm}\left( \theta \right) \frac{d\theta ^{l}\left( \tau \right) }{%
d\tau }\frac{d\theta ^{m}\left( \tau \right) }{d\tau }\right] ^{\frac{1}{2}%
}=1\text{,}  \label{nc1}
\end{equation}%
with $l=m=1$ and $g_{11}\left( \theta \right) =4\left( 1-\theta ^{2}\right)
^{-1}$. Equation (\ref{nc1}) implies that the temporal duration $\Delta \tau
^{\prime }$ needed to navigate the geodesic path connecting $\theta _{i}$ to 
$\theta _{f}$ is given by,%
\begin{equation}
\Delta \tau ^{\prime }\overset{\text{def}}{=}\tau _{f}^{\prime }-\tau
_{i}^{\prime }=2\left( \arcsin \theta _{f}-\arcsin \theta _{i}\right) \text{.%
}
\end{equation}%
We point out that $\Delta \tau ^{\prime }$ can also be viewed as the length $%
L\left( \theta _{i}\text{, }\theta _{f}\right) $ of the geodesic path
connecting $\theta _{i}$ to $\theta _{f}$,%
\begin{equation}
L_{\text{Model-II}}\left( \theta _{i}\text{, }\theta _{f}\right) \overset{%
\text{def}}{=}\int_{\theta _{i}}^{\theta _{f}}\left[ g_{lm}\left( \theta
\right) d\theta ^{l}d\theta ^{m}\right] ^{\frac{1}{2}}=\int_{\tau
_{i}}^{\tau _{f}^{\prime }}d\tau \left[ g_{lm}\left( \theta \right) \frac{%
d\theta ^{l}\left( \tau \right) }{d\tau }\frac{d\theta ^{m}\left( \tau
\right) }{d\tau }\right] ^{\frac{1}{2}}\text{.}
\end{equation}%
Indeed, in the case being considered $L_{\text{Model-II}}\left( \theta _{i}%
\text{, }\theta _{f}\right) $ reads%
\begin{equation}
L_{\text{Model-II}}\left( \theta _{i}\text{, }\theta _{f}\right) \overset{%
\text{def}}{=}\int_{\theta _{i}}^{\theta _{f}}\sqrt{\frac{4}{1-\theta ^{2}}}%
d\theta =2\left( \arcsin \theta _{f}-\arcsin \theta _{i}\right) \equiv
\Delta \tau ^{\prime }\text{.}
\end{equation}%
Comparing Grover's model to the deviation from Grover's model and assuming $%
\theta _{f}=\theta _{i}+\epsilon $ with $\theta _{i}\equiv 0$ and $%
0<\epsilon \ll 1$, we obtain%
\begin{equation}
\Delta \tau _{\text{Model-II}}-\Delta \tau _{\text{Grover}}\simeq \frac{%
\epsilon ^{3}}{3}+\mathcal{O}\left( \epsilon ^{5}\right) \geq 0\text{.}
\label{comparison}
\end{equation}%
We may conclude that navigating along Grover's geodesic path is more
efficient than navigating along the geodesic path characterizing the
modified statistical model. Indeed, assuming that any navigation that is
being compared occurs with the same unit speed, the efficiency of the
navigation is simply quantified in terms of the length of the geodesic path
connecting the initial and final states. Furthermore, since the length of a
geodesic path can be regarded as the navigation duration, the most efficient
unit-speed navigation is characterized by the shortest navigation duration
(shortest geodesic path) for a given set of boundary conditions.

\subsection{Deviation from Grover's Model: Parametric-Independence without
Actuality}

It is possible to consider normalized probability distribution vectors $\vec{%
p}_{\text{Model-III}}\equiv \left( p_{0}\left( \theta \right) \text{, }%
p_{1}\left( \theta \right) \text{,..., }p_{N-1}\left( \theta \right) \right) 
$ that satisfy the\textbf{\ }parametric-independence constraint but do not
minimize the action functional $\mathcal{S}^{\prime }\left[ q_{k}\left(
\theta \right) \right] $. For instance, assuming that $p_{k}\left( \theta
\right) =p_{k^{\prime }}\left( \theta \right) $ for $k$ and $k^{\prime }\in
\left\{ 1\text{,..., }N-1\right\} $, the components of $\vec{p}_{\text{%
Model-III}}$ should satisfy the following system of two equations,%
\begin{equation}
p_{0}\left( \theta \right) +\left( N-1\right) p_{\bar{k}}\left( \theta
\right) =1\text{ and, }\frac{\dot{p}_{0}^{2}\left( \theta \right) }{%
p_{0}\left( \theta \right) }+\left( N-1\right) \frac{\dot{p}_{\bar{k}%
}^{2}\left( \theta \right) }{p_{\bar{k}}\left( \theta \right) }=4\text{,}
\label{ed1}
\end{equation}%
with $\bar{k}=1$,..., $N-1$. From the first equation in (\ref{ed1}), we get%
\begin{equation}
p_{\bar{k}}\left( \theta \right) =\frac{1-p_{0}\left( \theta \right) }{N-1}%
\text{.}  \label{ed2}
\end{equation}%
Substituting (\ref{ed2})\ into the second equation in (\ref{ed1}), we obtain%
\begin{equation}
\dot{p}_{0}^{2}-4p_{0}\left( 1-p_{0}\right) =0\text{.}  \label{ed3}
\end{equation}%
Integrating (\ref{ed3})\ and using (\ref{ed2}), a solution of the system (%
\ref{ed1}) is given by,%
\begin{equation}
\vec{p}_{\text{Model-III}}\equiv \left( \frac{1+\sin 2\theta }{2}\text{, }%
\frac{1-\sin 2\theta }{2\left( N-1\right) }\text{,..., }\frac{1-\sin 2\theta 
}{2\left( N-1\right) }\right) \text{.}
\end{equation}%
In this case, the Fisher information function $\mathcal{F}\left( \theta
\right) =4$, however the components of $\vec{p}_{\text{Model-III}}$ do not
satisfy the EL equation in (\ref{EL}). Although the integration of the
geodesic equation in (\ref{GE}) for this model would lead to geodesic paths
that are straight line trajectories, the components of $\vec{p}_{\text{%
Model-III}}$ represent only possible and not actual trajectories of the
system (according to the principle of least action). Thus, the probability
path $\vec{p}_{\text{Model-III}}$ lacks the actuality requirement that
distinguishes Grover's probability path. Indeed, a part from an irrelevant
constant factor, it turns out that the probability distribution vector
related to Grover's algorithm is the only one satisfying both the
parametric-independence and minimization condition (actuality constraint)
simultaneously.

\subsection{Deviation from Grover's Model: Actuality without
Parametric-Independence}

For the sake of completeness, we mention one more possible scenario. There
might exist a probability distribution vector $\vec{p}_{\text{Model-IV}%
}\equiv \left( p_{0}\left( \theta \right) \text{, }p_{1}\left( \theta
\right) \text{,..., }p_{N-1}\left( \theta \right) \right) $ with $%
p_{k}\left( \theta \right) =$ $p_{k^{\prime }}\left( \theta \right) $ for $k$
and $k^{\prime }\in \left\{ 1\text{,..., }N-1\right\} $ that satisfies the
minimization without fulfilling the parametric-independence requirement. The
components of such a path should satisfy the following system of \emph{%
nonlinear} ordinary differential equations,%
\begin{equation}
\left\{ 
\begin{array}{c}
\ddot{q}_{0}\left( \theta \right) -\frac{\mathcal{\dot{L}}\left( \theta
\right) }{\mathcal{L}\left( \theta \right) }\dot{q}_{0}\left( \theta \right)
+\frac{\mathcal{L}\left( \theta \right) }{2}q_{0}\left( \theta \right) =0%
\text{,} \\ 
\\ 
\ddot{q}_{_{\bar{k}}}\left( \theta \right) -\frac{\mathcal{\dot{L}}\left(
\theta \right) }{\mathcal{L}\left( \theta \right) }\dot{q}_{_{\bar{k}%
}}\left( \theta \right) +\frac{\mathcal{L}\left( \theta \right) }{2}q_{_{%
\bar{k}}}\left( \theta \right) =0\text{,}%
\end{array}%
\right.   \label{fed}
\end{equation}%
where $q_{0}^{2}\left( \theta \right) =p_{0}\left( \theta \right) $ and $%
q_{_{\bar{k}}}^{2}\left( \theta \right) =p_{_{\bar{k}}}\left( \theta \right) 
$ for $\bar{k}=1$,..., $N-1$ with the normalization condition%
\begin{equation}
q_{0}^{2}\left( \theta \right) +\left( N-1\right) q_{_{\bar{k}}}^{2}\left(
\theta \right) =1\text{.}  \label{NC1}
\end{equation}%
The quantity $\mathcal{L}\left( \theta \right) $ in (\ref{fed}) is defined
as,%
\begin{equation}
\mathcal{L}\left( \theta \right) \overset{\text{def}}{=}2\left[ \dot{q}%
_{0}^{2}\left( \theta \right) +\left( N-1\right) \dot{q}_{_{\bar{k}%
}}^{2}\left( \theta \right) \right] ^{\frac{1}{2}}\neq \text{constant.}
\end{equation}%
Using the normalization condition (\ref{NC1}), the expressions for $\mathcal{%
L}\left( \theta \right) $ and $\mathcal{\dot{L}}\left( \theta \right) 
\overset{\text{def}}{=}\frac{d\mathcal{L}\left( \theta \right) }{d\theta }$
read%
\begin{equation}
\mathcal{L}\left( \theta \right) =2\dot{q}_{0}\sqrt{\frac{\left( N-1\right)
\left( 1-q_{0}^{2}\right) +q_{0}^{2}}{\left( N-1\right) \left(
1-q_{0}^{2}\right) }}  \label{l}
\end{equation}%
and%
\begin{equation}
\mathcal{\dot{L}}\left( \theta \right) =2\ddot{q}_{0}\sqrt{\frac{\left(
N-1\right) \left( 1-q_{0}^{2}\right) +q_{0}^{2}}{\left( N-1\right) \left(
1-q_{0}^{2}\right) }}+\frac{2q_{0}\dot{q}_{0}^{2}}{\left( N-1\right) \left(
1-q_{0}^{2}\right) ^{2}}\sqrt{\frac{\left( N-1\right) \left(
1-q_{0}^{2}\right) }{\left( N-1\right) \left( 1-q_{0}^{2}\right) +q_{0}^{2}}}%
\text{,}  \label{lp}
\end{equation}%
respectively. It is fairly clear that finding an analytical expression for
actual $N$-dimensional $\vec{p}_{\text{Model-IV}}$ in the absence of
parametric-independence may turn out to be quite nontrivial (for further
details, see Appendix B). It is interesting to notice that the same
computational challenge occurs when attempting to find analytical solutions
of a non-unitary quantum walk \cite{kendon2}. However,\textbf{\ }focusing
our attention on probability paths $\vec{p}$ with only two types of
components, $p_{0}$ and $p_{\bar{k}}$ with $\bar{k}=1$,..., $N-1$ and $N=2$,
we are able to show that the only possible probability distribution vector
satisfying the system (\ref{fed}) is Grover's path where $\mathcal{L}\left(
\theta \right) $ is constant (and equals two). In such scenario, Eqs. (\ref%
{l}) and (\ref{lp}) become%
\begin{equation}
\mathcal{L}\left( \theta \right) =\frac{2\dot{q}_{0}}{\sqrt{1-q_{0}^{2}}}%
\text{ and, }\mathcal{\dot{L}}\left( \theta \right) =\frac{2\ddot{q}%
_{0}\left( 1-q_{0}^{2}\right) +2q_{0}\dot{q}_{0}^{2}}{\left(
1-q_{0}^{2}\right) ^{\frac{3}{2}}}\text{.}  \label{questa}
\end{equation}%
Using (\ref{questa}) together with the first equation of (\ref{fed}), we get%
\begin{equation}
\dot{q}_{0}-\sqrt{1-q_{0}^{2}}=0\text{,}  \label{minor}
\end{equation}%
that is,%
\begin{equation}
q_{0}\left( \theta \right) =\sin \left( \theta +c_{0}\right) \text{ and }%
q_{1}\left( \theta \right) =\cos \left( \theta +c_{0}\right)   \label{minor1}
\end{equation}%
where $c_{0}$ is the \emph{real }integration constant. We then conclude that
in this special toy-case (one-qubit quantum Hilbert space, $\mathcal{H}%
_{2}^{1}$ with $\dim _{%
\mathbb{C}
}\mathcal{H}_{2}^{1}=2=N$), actual paths are necessarily
parametric-independent unitary paths as well. However, the realistic
scenario implies a continuous approximation of Grover's quantum searching
problem where $N\gg 1$. In such a case, actual parametric-dependent
probability paths with the two-component structure may still exist (see
Appendix B).

As a final remark, we point out that our information geometric analysis may
be of use to investigate the possibility of the existence of actual
parametric-dependent probability paths with a multi-component structure
(more that two types of components). Such scenarios occur in a multi-item
quantum search where more than a single item is being searched.

\section{Concluding Remarks and Open Issues}

In this article, we have quantified the notion of quantum distinguishability
between parametric density operators by means of the Wigner-Yanase quantum
information metric. We then presented an information geometric
characterization of Grover's quantum search algorithm as a geodesic in the
parameter space characterizing the pure state model, the manifold of the
parametric density operators of pure quantum states constructed from the
continuous approximation of the parametric quantum output state in Grover's
algorithm. Finally, we discussed few possible deviations from Grover's
algorithm within this quantum information geometric framework. For instance,
in the second example in Section V, we have shown that navigating along
Grover's geodesic path is more efficient than navigating along the geodesic
path characterizing the modified statistical model. This statement holds
when it is assumed that any navigation that is being compared occurs with
the same unit speed and, therefore, the efficiency of the navigation is
simply quantified in terms of the length of the geodesic path connecting the
initial and final states. Furthermore, since the length of a geodesic path
can be regarded as the navigation duration, we have also pointed out that
the most efficient unit-speed navigation is characterized by the shortest
navigation duration (shortest geodesic path) for a given set of boundary
conditions.

The quantum information geometric techniques employed in our work were
fairly simple. Unfortunately, we stress that there exist some technical
difficulties in handling quantum information geometric techniques in general
cases. For instance, there is no general formula for geodesic paths of
manifolds of density operators endowed with an arbitrary given monotone
Riemannian metric \cite{jencova}. Explicit expressions are known only for
two special metrics: the Bures metric \cite{dittmann} and the Wigner-Yanase
metric \cite{gibilisco}. In \cite{dittmann}, geodesic paths on manifolds of
density operators are obtained as projections of large circles on a large
sphere within the purifying Hilbert-Schmidt space. In \cite{gibilisco},
geodesic paths are obtained exploiting the classical pull-back approach to
Fisher information metric via sphere geometry. If diagonalization procedures
are used, it is possible to obtain an explicit expression for the Bures
distance of two arbitrary density matrices in the two-dimensional case \cite%
{hubner}. In the general $n$-dimensional case this is unfortunately not
possible. Thus, one of the greatest mathematical problems to face in order
to quantify the metric properties of arbitrary mixed states is that of the
explicit computation of eigenvalues of arbitrary Hermitian density matrices.
The scenario becomes even more difficult if we recall that in order to find
the geodesics one needs to integrate the geodesic equations expressed in
terms of the metric and the Christoffel connection coefficients (which are
expressed in terms of the first order changes in the metric). Furthermore,
to integrate the Jacobi-Levi-Civita equation of geodesic spread, one also
needs the explicit expression of the Riemannian curvature tensor (which
encodes information about the second order changes in the metric tensor on
the manifold of density operators). It may be that the first and second
order perturbation theory of linear operators may turn out to be useful in
estimating the changes in the metric tensor. However, it appears that thus
far this line of investigation has not provided useful results \cite{hubner}%
. There have been other attempts to compute in finite dimensions explicit
formulae for the Bures metric where no diagonalization procedure is
employed. These approaches only make use of the theory of matrix equations,
determinants and traces \cite{dittmann2}. Needless to say that further
research is needed to fully handle the quantum information geometric
formalism.

As pointed out in \cite{amari}, quantum information geometry is only in its
infancy, and much more research awaits to be performed. However, from the
preliminary results obtained in this article, we have reason to think that
quantum information geometric techniques may turn out to be especially
useful for describing and understanding the efficiency of both unitary and
non-unitary quantum walks used to implement quantum search algorithms in
quantum computing \cite{shenvi}. In particular, our analysis opens up new
lines of investigation that may deserve some attention. For instance, it
would be worth deepening our analysis and attempting to understand the
connection between computational complexity classes of quantum algorithms 
\cite{nielsen-book} and the complexity of the quantum geodesic paths
associated with them. These investigations are left for future works.

\begin{acknowledgments}
C. C. thanks Marco Lucamarini and David Vitali for useful discussions. The
research leading to these results has received funding from the European
Commission's Seventh Framework Programme (FP7/2007--2013) under grant
agreements no. 213681.
\end{acknowledgments}

\appendix

\section{Derivation of the Wigner-Yanase line element}

Here we compute in an explicit manner the infinitesimal Wigner-Yanase line
element $ds_{\text{WY}}^{2}$,%
\begin{equation}
ds_{\text{WY}}^{2}=4ds_{\text{FS}}^{2}\text{,}
\end{equation}%
where the Fubini-Study infinitesimal line element $ds_{\text{FS}}^{2}$ is
given by,%
\begin{equation}
ds_{\text{FS}}^{2}=\left\Vert d\psi \right\Vert ^{2}-\left\vert \left\langle
\psi |d\psi \right\rangle \right\vert ^{2}=1-\left\vert \left\langle \psi
^{\prime }|\psi \right\rangle \right\vert ^{2}\text{,}
\end{equation}%
where $\left\vert \psi \right\rangle $ and $\left\vert \psi ^{\prime
}\right\rangle $ are neighboring normalized pure states expanded in an
orthonormal basis $\left\{ \left\vert k\right\rangle \right\} $ with $k\in
\left\{ 1\text{,..., }N\right\} $,%
\begin{equation}
\left\vert \psi \right\rangle =\sum_{k=1}^{N}\sqrt{p_{k}\left( \theta
\right) }e^{i\phi _{k}\left( \theta \right) }\left\vert k\right\rangle \text{
and, }\left\vert \psi ^{\prime }\right\rangle =\sum_{k=1}^{N}\sqrt{%
p_{k}+dp_{k}}e^{i\left( \phi _{k}+d\phi _{k}\right) }\left\vert
k\right\rangle \text{,}
\end{equation}%
respectively. Observe that,%
\begin{eqnarray}
\left\vert \psi ^{\prime }\right\rangle  &=&\sum_{k=1}^{N}\sqrt{p_{k}+dp_{k}}%
e^{i\left( \phi _{k}+d\phi _{k}\right) }\left\vert k\right\rangle
=\sum_{k=1}^{N}\left[ \sqrt{p_{k}}\sqrt{1+\frac{dp_{k}}{p_{k}}}e^{i\phi
_{k}}\left( 1+id\phi _{k}+\frac{1}{2}\left( id\phi _{k}\right) ^{2}\right) %
\right] \left\vert k\right\rangle   \notag \\
&&  \notag \\
&=&\sum_{k=1}^{N}\left[ \sqrt{p_{k}}\left( 1+\frac{1}{2}\frac{dp_{k}}{p_{k}}-%
\frac{1}{8}\frac{dp_{k}^{2}}{p_{k}^{2}}\right) e^{i\phi _{k}}\left( 1+id\phi
_{k}-\frac{1}{2}d\phi _{k}^{2}\right) \right] \left\vert k\right\rangle 
\text{.}  \label{ff}
\end{eqnarray}%
Eq. (\ref{ff}) implies that,%
\begin{equation}
\left\langle \psi ^{\prime }\right\vert =\sum_{k=1}^{N}\left[ \sqrt{p_{k}}%
\left( 1+\frac{1}{2}\frac{dp_{k}}{p_{k}}-\frac{1}{8}\frac{dp_{k}^{2}}{%
p_{k}^{2}}\right) e^{-i\phi _{k}}\left( 1-id\phi _{k}-\frac{1}{2}d\phi
_{k}^{2}\right) \right] \left\langle k\right\vert \text{,}
\end{equation}%
and thus $\left\langle \psi ^{\prime }|\psi \right\rangle $ becomes,%
\begin{eqnarray}
\left\langle \psi ^{\prime }|\psi \right\rangle  &=&\sum_{k=1}^{N}\left[
p_{k}\left( 1+\frac{1}{2}\frac{dp_{k}}{p_{k}}-\frac{1}{8}\frac{dp_{k}^{2}}{%
p_{k}^{2}}\right) \left( 1-id\phi _{k}-\frac{1}{2}d\phi _{k}^{2}\right) %
\right]   \notag \\
&&  \notag \\
&=&\sum_{k=1}^{N}\left[ \left( p_{k}+\frac{1}{2}dp_{k}-\frac{1}{8}\frac{%
dp_{k}^{2}}{p_{k}}\right) \left( 1-id\phi _{k}-\frac{1}{2}d\phi
_{k}^{2}\right) \right]   \notag \\
&&  \notag \\
&=&\sum_{k=1}^{N}\left( p_{k}+\frac{1}{2}dp_{k}-\frac{1}{8}\frac{dp_{k}^{2}}{%
p_{k}}\right) -i\sum_{k=1}^{N}\left( p_{k}+\frac{1}{2}dp_{k}-\frac{1}{8}%
\frac{dp_{k}^{2}}{p_{k}}\right) d\phi _{k}-\frac{1}{2}\sum_{k=1}^{N}p_{k}d%
\phi _{k}^{2}  \notag \\
&&  \notag \\
&=&1-\frac{1}{8}\sum_{k=1}^{N}\frac{dp_{k}^{2}}{p_{k}}-i\sum_{k=1}^{N}p_{k}d%
\phi _{k}-\frac{i}{2}\sum_{k=1}^{N}dp_{k}d\phi _{k}-\frac{1}{2}%
\sum_{k=1}^{N}p_{k}d\phi _{k}^{2}\text{,}
\end{eqnarray}%
where we have made use of the normalization constraint and its differential
form,%
\begin{equation}
\sum_{k=1}^{N}p_{k}=1\text{ and, }\sum_{k=1}^{N}dp_{k}=0\text{,}
\end{equation}%
respectively. We then obtain that $\left\vert \left\langle \psi ^{\prime
}|\psi \right\rangle \right\vert ^{2}$ reads,%
\begin{eqnarray}
\left\vert \left\langle \psi ^{\prime }|\psi \right\rangle \right\vert ^{2}
&=&\left\langle \psi ^{\prime }|\psi \right\rangle \left\langle \psi
^{\prime }|\psi \right\rangle ^{\ast }  \notag \\
&&  \notag \\
&=&\left[ 1-\frac{1}{8}\sum_{k=1}^{N}\frac{dp_{k}^{2}}{p_{k}}%
-i\sum_{k=1}^{N}p_{k}d\phi _{k}-\frac{i}{2}\sum_{k=1}^{N}dp_{k}d\phi _{k}-%
\frac{1}{2}\sum_{k=1}^{N}p_{k}d\phi _{k}^{2}\right] \cdot   \notag \\
&&  \notag \\
&&\cdot \left[ 1-\frac{1}{8}\sum_{k=1}^{N}\frac{dp_{k}^{2}}{p_{k}}%
+i\sum_{k=1}^{N}p_{k}d\phi _{k}+\frac{i}{2}\sum_{k=1}^{N}dp_{k}d\phi _{k}-%
\frac{1}{2}\sum_{k=1}^{N}p_{k}d\phi _{k}^{2}\right]   \notag \\
&&  \notag \\
&=&1-\frac{1}{8}\sum_{k=1}^{N}\frac{dp_{k}^{2}}{p_{k}}+i\sum_{k=1}^{N}p_{k}d%
\phi _{k}+\frac{i}{2}\sum_{k=1}^{N}dp_{k}d\phi _{k}-\frac{1}{2}%
\sum_{k=1}^{N}p_{k}d\phi _{k}^{2}-\frac{1}{8}\sum_{k=1}^{N}\frac{dp_{k}^{2}}{%
p_{k}}+  \notag \\
&&  \notag \\
&&-i\sum_{k=1}^{N}p_{k}d\phi _{k}+\left( \sum_{k=1}^{N}p_{k}d\phi
_{k}\right) ^{2}-\frac{i}{2}\sum_{k=1}^{N}dp_{k}d\phi _{k}-\frac{1}{2}%
\sum_{k=1}^{N}p_{k}d\phi _{k}^{2}  \notag \\
&&  \notag \\
&=&1-\frac{1}{4}\sum_{k=1}^{N}\frac{dp_{k}^{2}}{p_{k}}-\sum_{k=1}^{N}p_{k}d%
\phi _{k}^{2}+\left( \sum_{k=1}^{N}p_{k}d\phi _{k}\right) ^{2}\text{.}
\end{eqnarray}%
In conclusion, the infinitesimal Wigner-Yanase line element $ds_{\text{WY}%
}^{2}$ becomes%
\begin{eqnarray}
ds_{\text{WY}}^{2} &=&4\left\{ \frac{1}{4}\sum_{k=1}^{N}\frac{dp_{k}^{2}}{%
p_{k}}+\left[ \sum_{k=1}^{N}p_{k}d\phi _{k}^{2}-\left(
\sum_{k=1}^{N}p_{k}d\phi _{k}\right) ^{2}\right] \right\}   \notag \\
&&  \notag \\
&=&\sum_{k=1}^{N}\frac{dp_{k}^{2}}{p_{k}}+4\left[ \sum_{k=1}^{N}p_{k}d\phi
_{k}^{2}-\left( \sum_{k=1}^{N}p_{k}d\phi _{k}\right) ^{2}\right]   \notag \\
&&  \notag \\
&=&\left\{ \sum_{k=1}^{N}\frac{1}{p_{k}}\left( \frac{dp_{k}}{d\theta }%
\right) ^{2}+4\left[ \sum_{k=1}^{N}p_{k}\left( \frac{d\phi _{k}}{d\theta }%
\right) ^{2}-\left( \sum_{k=1}^{N}p_{k}\frac{d\phi _{k}}{d\theta }\right)
^{2}\right] \right\} d\theta ^{2}  \notag \\
&&  \notag \\
&=&\left\{ \sum_{k=1}^{N}\frac{\dot{p}_{k}^{2}}{p_{k}}+4\left[
\sum_{k=1}^{N}p_{k}\dot{\phi}_{k}^{2}-\left( \sum_{k=1}^{N}p_{k}\dot{\phi}%
_{k}\right) ^{2}\right] \right\} d\theta ^{2}\text{,}
\end{eqnarray}%
that is,%
\begin{equation}
ds_{\text{WY}}^{2}=\left\{ \sum_{k=1}^{N}\frac{\dot{p}_{k}^{2}}{p_{k}}+4%
\left[ \sum_{k=1}^{N}p_{k}\dot{\phi}_{k}^{2}-\left( \sum_{k=1}^{N}p_{k}\dot{%
\phi}_{k}\right) ^{2}\right] \right\} d\theta ^{2}\text{,}
\end{equation}%
where,%
\begin{equation}
\dot{p}_{k}=\frac{dp_{k}\left( \theta \right) }{d\theta }\text{ and, }\dot{%
\phi}_{k}=\frac{d\phi _{k}\left( \theta \right) }{d\theta }\text{.}
\end{equation}

\section{Actual Parametric-Dependent Unitary Probability Distribution Vectors%
}

Substituting (\ref{l}) and (\ref{lp}) into the first equation in (\ref{fed}%
), the EL equation reads%
\begin{equation}
\frac{dq_{0}\left( \theta \right) }{d\theta }-\frac{1}{\sqrt{N-1}}\left(
1-q_{0}^{2}\right) ^{\frac{1}{2}}\left[ \left( N-1\right) \left(
1-q_{0}^{2}\right) +q_{0}^{2}\right] ^{\frac{3}{2}}=0\text{,}
\end{equation}%
that is,%
\begin{equation}
\mathcal{I}_{N}\left( q_{0}\left( \theta \right) \right) \overset{\text{def}}%
{=}\int \frac{dq_{0}}{\left( 1-q_{0}^{2}\right) ^{\frac{1}{2}}\left[ \left(
N-1\right) \left( 1-q_{0}^{2}\right) +q_{0}^{2}\right] ^{\frac{3}{2}}}=\frac{%
1}{\sqrt{N-1}}\int d\theta \text{.}  \label{int}
\end{equation}%
The integral $\mathcal{I}\left( q_{0}\left( \theta \right) \right) $ on the
left-hand side of (\ref{int}) can be expressed in terms of an elliptic
integral of the second kind $E\left[ \cdot |\cdot \right] $ (see \cite%
{stegun}, for instance) as follows,%
\begin{equation}
\mathcal{I}_{N}\left( q_{0}\left( \theta \right) \right) =\frac{\sqrt{%
N\left( 1-q_{0}^{2}\right) +2q_{0}^{2}-1}}{\left( N-1\right) }\cdot \left[ 
\frac{\left( N-2\right) q_{0}\sqrt{1-q_{0}^{2}}}{N\left( q_{0}^{2}-1\right)
-2q_{0}^{2}+1}+\frac{E\left( \sin ^{-1}\left( q_{0}\right) |\frac{N-2}{N-1}%
\right) }{\sqrt{\frac{N\left( 1-q_{0}^{2}\right) +2q_{0}^{2}-1}{N-1}}}\right]
\text{.}
\end{equation}%
Finding an analytical solution $q_{0}\left( \theta \right) $ satisfying the
following equation,%
\begin{equation}
\mathcal{I}_{N}\left( q_{0}\left( \theta \right) \right) -\left[ \frac{1}{%
\sqrt{N-1}}\theta +C_{N}\right] =0\text{,}  \label{diff}
\end{equation}%
where $C_{N}$ is a \emph{real} integration constant is quite challenging.
Approximate solutions to Eq. (\ref{diff}) would require numerical
investigations which are left for elsewhere. However, in the two-dimensional
case ($N=2$), we get a closed form solution for $q_{0}\left( \theta \right) $%
. In such a case $\mathcal{I}_{2}\left( q_{0}\left( \theta \right) \right) $
reads,%
\begin{equation}
\mathcal{I}_{2}\left( q_{0}\left( \theta \right) \right) =E\left( \sin
^{-1}\left( q_{0}\right) |0\right) \overset{\text{def}}{=}\arcsin \left(
q_{0}\left( \theta \right) \right) \text{,}
\end{equation}%
that is,%
\begin{equation}
q_{0}\left( \theta \right) =\sin \left( \theta +C_{2}\right) \text{.}
\label{sole}
\end{equation}%
As a conclusive remark, we stress that although finding an analytical
solution $q_{0}\left( \theta \right) $\ for Eq. (\ref{diff}) may not be
trivial, for sure $q_{0}\left( \theta \right) $\ in (\ref{sole}) does not
satisfy this equation when $N\neq 2$. This leaves open the possibility of
finding actual parametric-dependent unitary probability paths also in the
case of a single-item quantum search.

\end{document}